\documentclass[12pt,preprint]{aastex}

\shorttitle{Spherical Collapse in Modified Newtonian Dynamics
(MOND)} \shortauthors{M. Malekjani, S. Rahvar \&H. Haghi}
\begin{document}

\title{Spherical Collapse in Modified Newtonian Dynamics (MOND)}

\author{M. Malekjani\altaffilmark{1}, S. Rahvar
\altaffilmark{2}, H. Haghi \altaffilmark{3,4}}
\email{malekjani@tabrizu.ac.ir} \email{rahvar@sharif.edu}
\email{haghi@mehr.sharif.edu}

\altaffiltext{1}{Department of Theoretical Physics and Astrophysics,
University of Tabriz, P.O.Box 51664 Tabriz, Iran}

\altaffiltext{2}{Department of Physics, Sharif University of
Technology, P.O.Box 11365--9161, Tehran, Iran}

\altaffiltext{3}{ Institute for Advanced Studies in Basic Sciences,
P. O. Box 45195-1159, Zanjan, Iran}

\altaffiltext{4}{Institute for Studies in Theoretical Physics and
Mathematics, P.O.Box 19395--5531, Tehran, Iran}

\begin{abstract}
Modeling the structure formation in the universe, we extend the
spherical collapse model in the context of MOND starting with the
linear Newtonian structure formation followed by the MONDian
evolution. In MOND the formation of structures speed up without a
need for dark matter. Starting with the top-hat over-dense
distribution of the matter, the structures virialize with a
power--law profile of the distribution of matter. We show that the
virialization process takes place gradually from the center of the
structure to the outer layers. In this scenario the smaller
structures enter to the MONDian regime earlier and evolve faster,
hence they are older than larger structures. We also show that the
virialization of the structures occur in the MONDian regime, in
which the smaller structures have stronger gravitational
acceleration than the larger ones. This feature of the dynamical
behavior of the structures is in agreement with this fact that the
smaller structures as the globular clusters or galactic bulges have
been formed earlier and need less dark matter in CDM scenario.
\end{abstract}

\section{Introduction}
The conjecture for the existence of dark matter dates back to Zwicky
in 1933, who failed to explain the dynamics of Coma cluster by the
virial theorem through the distribution of visible matter
\cite{zwi}. To interpret the dynamics of the structure, the concept
of missing mass or dark matter entered the Astrophysical studies
since then. In addition to the cosmological scales, in the smaller
galactic scales, the flat rotation curve of the spiral galaxies
requires the existence of the dark matter halo \cite{bos81}. On the
other hand studying the dynamics of the universe as a whole reveals
that the universe is dominated by the dark matter and dark energy
\cite{spe03}. This hypothetical matter neither emits light nor
interacts with the ordinary matter and only shows its presence
through its gravitational interaction. The advantage of the Cold
Dark Matter ($CDM$) model is that, it can successfully explain the
rotation curve of the spiral galaxies and lensing by galaxies and
cluster of galaxies. Within the framework of general relativity (GR)
it also provides a reasonable description for the hierarchy in the
structure formation. Although currently $CDM$ and $\Lambda CDM$
models are remarkably successful in large scales \cite{spe03}, they
cannot explain Tully-Fisher and Freeman laws \citep{bosch00}. On the
other hand, the high-resolution N-body simulations are still in
contradiction with the observations on sub-galactic scales where
they predict orders of magnitude more substructures than what is
observed \citep{moo99,kly99}. They also provide incompatible spatial
distribution of the sub-halos \cite{kro04}.

The other alternative to the dark matter is the modification to the
gravity law by means of taking a generic form of action for the
gravity rather than that of the Einstein-Hilbert action. This
approach has been introduced to be an alternative model to the dark
energy \cite{car04}. $f(R)$ gravity also is used to interpret the
rotation curve of the spiral galaxies \cite{sob07,saf08}. There is
also recent efforts on modifying the gravity law  by using a simple
kinetic Lagrangian whether the pressure can bend space-time
sufficiently to replace the roles of dark energy, cold dark matter
and heavy neutrinos in explaining anomalous accelerations at all
scales \cite{zhao07}. Halle et al (2008) also proposed a generalized
lagrangian with non-uniform cosmological constant for the vacuum
field within the framework of the Einstein gravity.

Finally the third approach, which we are concerned with in this work
is the modification of the conventional Newtonian law is so-called
MOdified Newtonian Dynamics (MOND). The dynamics of a structure in
MOND under the gravitational field is given by \cite{mil83}:
\begin{equation}
\mu(g/a_0)\bf{g} = \bf{g_N}, \label{mond}
\end{equation}
where $g_N$ is the Newtonian gravitational acceleration, $a_0
=1.2\times 10^{-10}ms^{-2}$ is a fundamental acceleration parameter
and $\mu(x)$ is a function for the transition from the Newtonian to
the MONDian regime (e.g. $\mu(x) = x/\sqrt{1+x^2}$). The dynamics of
the structure for $g<a_0$ deviates from the Newtonian law by
$\mu(x)$ and for $a_0\ll g $  we recover the Newtonian dynamics
(i.e. $\mu =1$). On the other hand for $g \ll a_{0}$, so-called deep
MOND regime, $\mu(x) = x$
 and the effective acceleration is given
by $g=\sqrt{g_{N}a_{0}}$.

Due to confusion in the definition of the dynamical concepts in
MOND, this model can be interpreted as a modification to the gravity
law instead of dynamics. Bekenestein and Milgrom (1984) used a
non-conventional Newtonian action for the gravity to extract the
modified Poisson equation as follows:
\begin{equation}
\nabla.\left[\mu(\nabla\phi/a_0)\bf{\nabla\phi}\right] = 4\pi G
\rho, \label{mond2}
\end{equation}
where in the spherical symmetric systems this equation reduces to
equation (\ref{mond}). One of the problems with MOND is that it is
not a covariant gravity model. Bekenestein (2004) proposed a
covariant formulation of this model. This theory in addition to the
metric has a scalar field as well as a 4-vector field called TeVeS,
where in the limit of non-relativistic, small accelerations, the
field equation reduces to that in MOND. The total potential in this
theory is given by the sum of the Newtonian potential $\phi_N$ and
the potential due to the scalar field, $\phi_s$:
\begin{equation}
 \Phi = \Phi_{N} + \phi_s ,
\end{equation}
where the added scalar field plays the role of the dark matter.
TeVeS theory has also a Newtonian limit for the non-relativistic
dynamics with significant acceleration. In the non-relativistic
limit we can simplify the gravity inside the spherical system as
\begin{equation}
g(r) = -\nabla\Phi \simeq \left\{
                           \begin{array}{ll}
                             \sqrt{a_0 g_N}, & g_N \leqslant a_0; \\
                             g_N, & otherwise.
                           \end{array}
                         \right.
\label{eqn:eqmotionTeves2}
\end{equation}
The advantage of MOND is that it could provide a successful fit to
the rotation curve of spiral galaxies and dispersion velocity of
elliptical galaxies \citep{mil03,san96}. It has been tested against
the Cosmic Background Radiation \cite{sko06}, gravitational lensing
\cite{che06,zha06,ang07}, stellar systems and galactic dynamics
\citep{hagh06,lon07,tir07}, solar system \cite{bek06,ser06} and
Tully-Fisher and Freeman laws \cite{Mcg99,Mcg00}. MOND also
decreases the mass discrepancy in the cluster of galaxies
\cite{poi05} but yet in clusters it remains necessary to invoke the
undetected matter, possibly in the form of a massive neutrino
\cite{san99,san03,agu02,san02}.

Studying the cosmology and formation of the large structures in the
universe is another tool to examine MOND. Studying the MONDian
scenario of the structure formation has been started by Felten
(1984) and Sanders (1998). In the paper by Sanders (1998), it is
shown that a patch of universe in the MONDian regime smaller than
the horizon size can evolve with a different rate than the
background, hence the structures naturally can be formed through
this scale dependent dynamics. In this scenario small structures
form before the larger ones and this provides a bottom-up
hierarchical procedure for the formation of the structures in the
universe. The problem with this model is that the center of collapse
is not identified and every point in the space depending on the
 choice of the coordinate system can be considered as the center of the structure.
For solving this problem one can consider MOND formula to be applied
to a peculiar acceleration developing from density fluctuation
rather than Hubble expansion \cite{san01,nus01}. Nusser (2002) used
the amplitude of the CMB anisotropies as the initial condition in
the N-body simulation to simulate the large scale structures in the
universe. However to have a compatible result, one has to either
decrease $a_0$ by one order of magnitude or reduce the amplitude of
the fluctuations at the initial condition. Recent studies in the
formation of the galaxies by Sanders (2008) shows that massive
elliptical galaxies may be formed at $z>10$, as a consequence of the
monolithic dissipation-less collapse. Applying MOND with cooling
mechanism put an upper limit to the stellar clustering in the form
of the galaxy. Extending the structure formation in TeVeS theory has
been done by linear perturbation of metric, vector and scalar
fields. The predictions are compatible with the observations of the
structures \cite{dod}. Skordis (2008) also used a generalized TeVeS
theory to construct the primordial adiabatic perturbations of a
general family of the scalar field kinetic functions.

In this work we extend the spherical collapse model in MOND for
studying the general behavior of the structures during their
formation. The initial baryonic density contrast for the structures
is taken from the CMB anisotropy. We obtain the dynamics of the
baryonic structures in the early epoches with the linear Newtonian
structure formation until the entering of the structure to the
MONDian regime and follow the evolution with the MOND. The dynamical
evolution with MOND shows that the structures virialize with
power-law distribution of matter. We show that while all of the
structures re-collapse and virialize in the MONDian regime, the
gravitational acceleration of the structures in this stage inversely
depends on the size of the structure. This dynamical behavior of the
structure formation is compatible with the less-existence of the
dark matter in the globular clusters and central parts of the
galaxies in CDM scenario.

The organization of the paper is as follows: In section \ref{str} we
give a brief review of the spherical collapse model in MOND and
extend it by looking at the dynamics of each layer in the onion
model. In section \ref{evol} we discuss about the results of the
calculation, showing that the density profile inside the structure
is a time varying function during the evolution and the structure at
the final stage, virialize from the center to the outer areas. In
section \ref{conc} we summarize and discuss the results.
\section{Structure Formation in MOND}
\label{str} In this section we introduce the MONDian cosmology and
apply MOND in the spherical collapse model for studying the
formation of the structures. Here we take the onion model for the
spherical structure, dividing the sphere to the co-centric shells
and studying the evolution of each layer separately and the
structure as a whole.
\subsection{MONDian cosmology}\label{initial}
\label{mondcos} In the standard cosmology the dynamics of the
universe in the matter dominated regime can be derived from the
Newtonian gravity. Sanders (1998) used this approach to obtain the
dynamics of universe in the MONDian scenario. In MODian cosmology, a
patch of universe can evolve with a different rate than the
background as soon as the acceleration fulfills the condition of
$g<a_0$. The result is dynamical decoupling of the smaller scales
from the background which causes the production of the over dense
regions in the universe. The reason for this feature of cosmological
dynamics in MOND is that unlike the Newtonian mechanics, the
acceleration in the comoving frame depends on the length scale.

Let us take a spherical region with radius $r$ from the background
in which $a_0 \ll g(r)$. Using the Newtonian dynamics the
acceleration is given by
\begin{equation}
\ddot{r}=-\frac{GM}{r^{2}}, \label{Newt_equation}
\end{equation}
where $M$ is the active gravitational mass and to have a compatible
relation with the relativistic results, we define it to be composed
of the relativistic and non-relativistic matter as
\begin{equation}
M=\frac{4\pi r^{3}}{3}(\rho+3p), \label{mass}
\end{equation}
where $\rho$ and $p$ are the density and pressure of the cosmic
fluid. Substituting equation (\ref{mass}) in equation
(\ref{Newt_equation}), the acceleration is given by:
\begin{equation}
\ddot{r}=-\frac{4\pi G}{3}(\rho + 3p) r. \label{nc}
\end{equation}
In equation (\ref{nc}), the gravitational acceleration increases
linearly with $r$. This implies that there should be a critical
radius $r_{c}$ where inside it the acceleration is smaller than the
MOND threshold $a_{0}$ and the dynamics is given by the MOND where
outside that radius it is Newtonian. This critical length scale
obtain by equaling the left hand side of equation (\ref{nc}) with
$\ddot{r} = - a_0$ as
\begin{equation}
r_{c}=\frac{3 a_0}{4\pi G(\rho+3p)}. \label{rc}
\end{equation}
This length scale separates the Newtonian $(r>r_c)$ and the MONDian
$(r<r_c)$ domains. As the density of universe changes with the
expansion of the universe, the critical radius also changes with
time. From the continuity equation, the matter and the radiation
densities vary as $\rho=\rho_0 a^{-3}$ and $p = p_0 a^{-4}$. Using
the definition of the critical density $\rho_c = 3 H_0^2/8 \pi G$,
equation (\ref{rc}) can be written in terms of the density
parameters, $\Omega$, and the scale factor
\begin{equation}
r_{c}=\frac{2a_{0}}{H_{0}^{2}|\Omega_{b}^{(0)}a^{-3}+2\Omega_{r}^{(0)}a^{-4}-2\Omega_{\Lambda}^{(0)}|},
\label{rc2}
\end{equation}
where we adapt $H_{0}=75 km s^{-1} Mpc^{-1}$,
$\Omega_{b}^{(0)}=0.02$, $\Omega_{r}^{(0)}=5\times10^{-5}$ and
$\Omega_{\Lambda}^{(0)}=0$.

Now we do comparison of the size of a structure with $r_c$. From
equation (\ref{rc2}) the critical radius $r_c$ changes with the
scale factor as $r_c\propto a^{4}$ in the radiation and $r_c\propto
a^{3}$ in the matter dominant epoches. On the other hand the size of
the structure is proportional to the scale factor (i.e. $\lambda
\propto a$). So we expect that the Newtonian structures eventually
will enter the MONDian regime as $r_c$ grows faster than $\lambda$.
Using the adapted cosmological parameters, the critical radius at
the present time is obtained $r_c\simeq 10 H_0^{-1}$ which means
that the whole observable universe resides in a MONDian domain.
Comparing these two length scales at the last scattering surface
results in $r_c/H^{-1}\simeq 10^{-4}$. The mass corresponding to
this critical radius at last scattering surface is about $M_{c}
\simeq 10^4 M_\odot$ and for $M<M_c$ the dynamics is given by MOND.
While we expect to have density contrast growth for the scales
$r<r_c$ at the decoupling, comparing the critical mass $M_c \simeq
10^4 M_\odot$ with the Jeans mass of $M_J \simeq 10^{5}M_\odot$ at
this time indicates that the structures at these scales should be
washed out by the pressure \cite{san98}.

The other feature of MONDian cosmology is that unlike the standard
cosmology where decoupling redshift is smaller than the equality
redshift ($z_{de}<z_{eq}$), in MOND the equality epoch is much after
than the decoupling time. This feature results from this fact that
decoupling is related to the baryonic density of the universe and
the temperature and both parameters depend only on the scale factor,
independent of the dynamics of the universe. So we expect to have
the same decoupling redshift in the MOND as the standard cosmology.
However since the dark matter does not exist in the MONDian
cosmology, the equality will be shifted to the lower redshifts. For
our adopted cosmological parameters the equality redshift is
obtained $z_{eq} = 400$.

\subsection{Structure Formation: Spherical Collapse Model}
\begin{figure}
\begin{center}
\includegraphics[angle=0,scale=0.7]{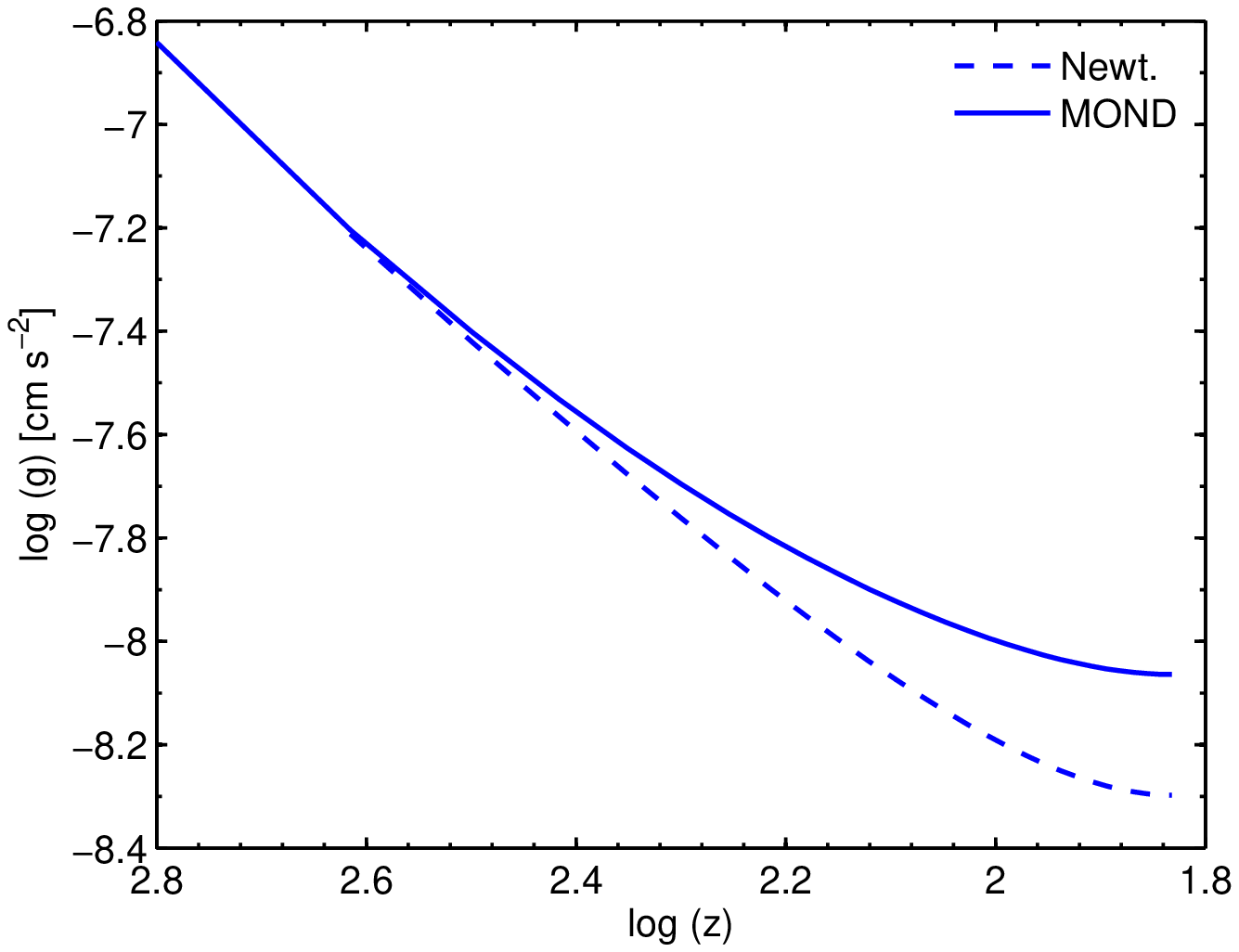}
\caption{Comparison of the accelerations in the Newtonian
(dashed-line) and MONDian (solid-line) regimes for a region with a
galaxy mass scale ($M=10^{11}M_\odot$) as a function of redshift.
For the early universe the difference between the two dynamics is
small but increases at the later times. The corresponding redshift
for entering of the structure to the critical radius is $z_{enter}
= 146$.} \label{N-MOND}
\end{center}
\end{figure}

In this part we model the evolution of a structure with an
over-dense spherical region in MOND. To calculate the evolution of
this spherical patch, for simplicity we take this over-dense region
with a top-hat distribution of matter. As the acceleration of this
structure depends on the distance from the center, we expect to have
different dynamics for each radius. Hence, we divide the structure
into the co-centric spherical shells like an onion model in
cosmology \cite{lem33,tol34,bon47} and calculate the dynamics of
each shell separately. The initial density contrast for this
over-dense region is taken from the fluctuations of the last
scattering surface. In the Newtonian treatment of the structure
formation $(r_c<\lambda)$, the density contrast grows linearly
$(\delta \propto a)$ from the decoupling epoch up to the entrance of
the structure to $r_c$. For the MONDian regime $(\lambda<r_c)$, we
switch the dynamics to MOND an calculate the evolution of the
structure. As an example let us take a sphere with a mass of
$M=10^{11}M_{\odot}$ and find its acceleration with respect to the
center. In Figure (\ref{N-MOND}) we compare the acceleration of this
spherical structure in MONDian and Newtonian dynamics as a function
of the redshift. We note that the redshift is defined according to
the dynamics of the scale factor at the background. At the early
epoches, the difference between these two dynamics is small as
$r_c<\lambda$, but after entering the structure to the MONDian
regime $\lambda \lesssim r_c$, the evolution of the structure by the
Newtonian dynamics and MOND start to diverge. This deviation of the
dynamics from that of Newtonian plays the role of the dark matter in
the standard scenario of the structure formation.

\begin{figure}
\begin{center}
\includegraphics[angle=0,scale=0.7]{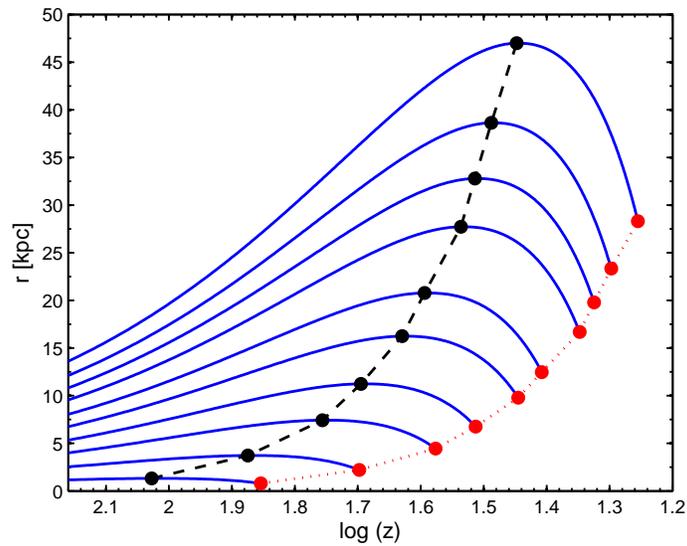}
\caption{The evolution of shells inside the spherical structure with
a galaxy mass $( 10^{11}M_{\odot})$ as a function of redshift, $r_i
= r_i(z)$. We take ten equidistant shells at the initial condition
for representing the evolution of each shell. The trend of the
evolution is reaching to a maximum radius, recollapsing and
virializing . The dashed line connects the maximum radii of the
shells and the dotted line indicates the virialized line, connecting
the virilize points of each shell.} \label{R-G}
\end{center}
\end{figure}

For calculating the dynamics of each shell in the onion model, we
take the following notation: $r^i(t)$ is the radius of $i$th shell
as a function of time and  $t^i_{ent}$ and $r^i_{ent}$ are the
entering time and radius of the $i$th shell to the MONDian domain,
respectively. The velocity of $i$th shell in terms of the Hubble
parameter at entrance time is given by $v^{i}_{ent}= H_{ent}
r^i_{ent}(1-\delta^i_{ent})$, where $H_{ent}$ is the Hubble
parameter of the Newtonian background and $\delta^i_{enter}$ is the
density contrast of the sphere enveloped by $r^i$. As we discussed
in the previous section, all the shells will eventually enter the
MONDian regime and we take this time as the initial condition for
each shell in the MONDian evolution of the structures. Table
(\ref{tab1}) shows the initial density contrast, radius and the
corresponding redshift of the entrance of each shell to the MONDian
regime. In the MONDian regime, the acceleration approximately is
$\sqrt{g_{N}a_0}$ and the evolution of each shell is as follows:
\begin{equation}
\ddot{r^i}=-\frac{\sqrt{GM^i a_0}}{r^i}, \label{monddyn}
\end{equation}
where $M^i$ is the mass of the structure enveloped by the $i{\it
th}$ shell. Using the initial conditions given by Table (1), we
obtain the evolution of each shell as shown in Figure (\ref{R-G}).
To visualize the evolution of the shells, we divide the sphere into
ten equidistant shells when all the structure is in the Newtonian
regime and obtain their evolution as a function of background
redshift. The dynamics of shells shows that the inner shells evolve
faster, reaching to a maximum radius and then collapse earlier than
the outer ones. Here the initial radius of outermost shell is about
$14 kpc$ at the entrance time of $z_{enter} \sim 146$ to the MOND
regime. This shell expands up to a maximum radius of $\sim49 kpc$ at
$z_{max}\sim 28$. Eventually, the shell starts to collapse and
virialize at $z\simeq 18$ with the radius of $\sim28.5 kpc$. The
maximum radius of each shell, $r_{max}$, is obtained from
integrating equation (\ref{monddyn}), letting $\dot{r}(t)=0$ as
follows:
\begin{equation}\label{alpha}
r^i_{max}=r^i_{ent}e^{\alpha}, ~~~
\alpha=\frac{{v^i_{ent}}^{2}}{2\sqrt{GM^i a_0}}.
\end{equation}
The next phase of the evolution of shells after reaching to a
maximum radius is re-collapsing. Similar to the standard scenario of
the spherical collapse models we expect that the global radial
velocity of the structure during the free fall collapse convert to
the dispersion velocity and prevent the structure from a
catastrophic collapse. This steady stage of the structure is given
by the virial theorem. The corresponding radius that fulfill the
virial condition is called the virial radius and is calculated from:
\begin{equation}\label{energy_cons}
\frac{1}{2}r^i\frac{d V(r^i)}{dr^i}+ V(r^i) = E, \label{vir}
\end{equation}
where $V(r^i)$ is the gravitational potential at the $i$th shell.
The potential in the Newtonian or MONDian regimes is given by
\begin{equation}
V(r^i) = \left\{
                           \begin{array}{ll}
                             \sqrt{GM^ia_0}\ln(r^i), & MOND; \\
                             -\frac{GM^i}{r^i}+\frac{GM^i}{r^i_{ent}} + \sqrt{GM^ia_0}\ln(r^i_{ent}). & Newt.
                           \end{array}
                         \right.
\label{pot2}
\end{equation}
For the non-dissipative evolution of the structure, the total energy
is conserved and we substitute the right hand side of equation
(\ref{vir}) by the energy of the system at the enterance time to the
MONDian regime, $E = \frac{1}{2}H_{ent}^2 {r^i_{ent}}^2+\sqrt{GM^i
a_0}\ln r^i_{ent}$. Using the potentials given by equation
(\ref{pot2}) at the left hand side of the equation (\ref{vir}), the
virial radius for the MOND and Newtonian regimes obtain as
\begin{equation}
r^i_{vir} = \left\{
                           \begin{array}{ll}
                             r^i_{ent}e^{\alpha-1/2}, & MOND; \\
                             r^i_{ent}(2-\frac{H_{ent}^2{r^i_{ent}}^2}{\frac{G M^i}{r^i_{ent}}})^{-1}. & Newt.
                           \end{array}
                         \right.
\label{vir2}
\end{equation}
If the virialization of the structure takes place in the MONDian
regime, $r^i_{vir}>r^i_{ent}$ implies $\alpha>1/2$ which results in
$2T^i_{ent}/W^i_{ent}>1$ where $T^i_{ent}$ is the kinetic energy for
a unit mass and $W^i_{ent} = r^i {dV^i}/{dr^i}$. On the other hand
if the structure virializes in the Newtonian regime, $
r^i_{vir}<r^i_{ent}$ condition from the equation (\ref{vir2})
implies $2T^i_{ent}/W^i_{ent}<1$. We calculate the expression of
$2T^i_{ent}/W^i_{ent}$ for each shell (see Table \ref{tab1}) and
show that all the shells of the structure virialize in the MONDian
regime (i.e. $r_{vir}^i>r_{ent}^i$).

Table (\ref{tab2}) shows the parameters of shells for the moment of
maximum radius and the virialization stage and Fig.(\ref{R-G})
visualizes quantitative behavior of the shells during their
evolution. In this figure the spots on the evolution curves shows
the two critical stages of the maximum radius and the virialization
radius, reported in Table (\ref{tab2}). The evolution lines of the
shells show that the inner shells evolve faster and virialize at the
higher redshifts, while the outer shells evolve slower.
\section{Predictions of the Model}
\label{evol} 
\begin{figure}
\begin{center}
\includegraphics[angle=0,scale=0.7]{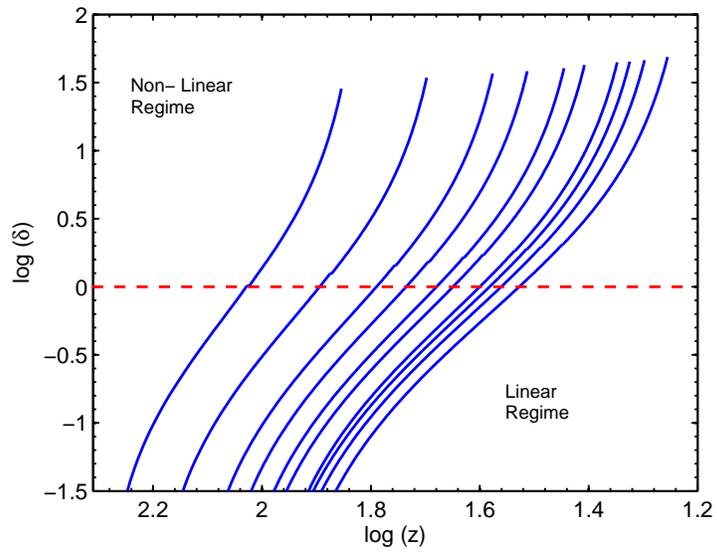}
\caption{A log-log plot of the density contrast evolution for each
shell as a function of the background redshift. Curves from the
left to the right corresponds the inner to the outer shells of the
structure. The horizon line represents $\delta=1$, separate the
linear and non-linear regimes. The inner layers reach to the
non-linear regime at the higher redshifts while the outer ones
become non-linear at the lower redshifts.}\label{dens}
\end{center}
\end{figure}

\begin{figure}
\begin{center}
\includegraphics[angle=0,scale=0.7]{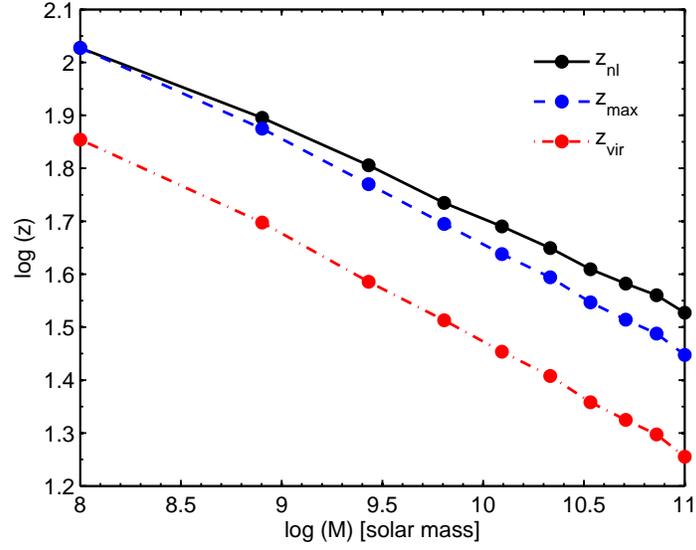}
\caption{Dependence of the non-linear redshift (solid line), maximum
radius redshift (dashed-line) and virialized redshift (dotted-dashed
line) as a function of the mass of structure in logarithmic scale.
Smaller structures enter to the non-linear regime at higher
redshifts and the larger ones enter at lower redshifts. This feature
of dependence of mass to the characteristic redshifts reveals that
the smaller structure in this model form before the larger ones.}
\label{f4}
\end{center}
\end{figure}

In this section we discuss the evolution of the density
contrast and the profile of matter distribution inside the
structure. We use the definition of the density contrast of the
shells in our model:
\begin{equation}
\delta^i(t)=\frac{\rho^i(t)-\bar\rho(t)}{\bar\rho(t)},
\end{equation}
where $\rho^i$ is the density of $i$th shell and $\bar\rho$ is the
density of the background. As the collapsing of the shells starts
from the inner to the outer parts of the structure, we will not have
a shell crossing during the evolution of the structure and from the
conservation of the mass, we can perform the Jacobian transformation
from the initial distribution of matter inside the sphere to the
evolved distribution as follows:
\begin{equation}
\rho^i=\rho^i_{ent}(\frac{r^i_{ent}}{r^i})^2\frac{\delta
r^i_{ent}}{\delta r^i}, \label{rho}
\end{equation}
where $\delta r^i$ represents the thickness of the $i{\it th}$
shell. Substituting the dynamics of each shell from the previous
section in equation (\ref{rho}), we obtain the evolution of density
contrast for the shells up to the virialization stage as shown in
Figure (\ref{dens}). The inner shells evolve faster and reach the
non-linear regime $(\delta>1)$ at the higher redshifts while the
outer ones evolve slower. Figure (\ref{f4}) shows the dependence of
the corresponding non-linear redshift to the mass of structure.
Comparing a small scale structure with the mass of $\sim 10^8
M_\odot$ with a galaxy having the mass of $\sim 10^{11} M_\odot$
shows that the former structure enters the non-linear regime at
$z_{nl}\simeq 106$ while the later one becomes non-linear at
$z_{nl}\simeq 33$. More details on the characteristic redshifts of
the structures in terms of their masses is reported in Table
(\ref{tab2}). From this table we extrapolate the dependence of the
non-linear redshift, maximum radius redshift and virialization
redshift to the mass of structure with the following functions (see
Fig. \ref{f4}):
\begin{eqnarray}
\log(z_{nl}) &=& -0.167 \log(\frac{M}{M_{\odot}}) + 3.38, \nonumber\\
\log(z_{max}) &=& -0.192\log(\frac{M}{M_{\odot}}) + 3.56, \nonumber\\
\log(z_{vir}) &=& -0.203 \log(\frac{M}{M_{\odot}}) + 3.48.
\end{eqnarray}
In what follows we describe the qualitative predictions of this
simple model.
\subsection{Density Profile}
\begin{figure}
\begin{center}
\includegraphics[angle=0,scale=0.7]{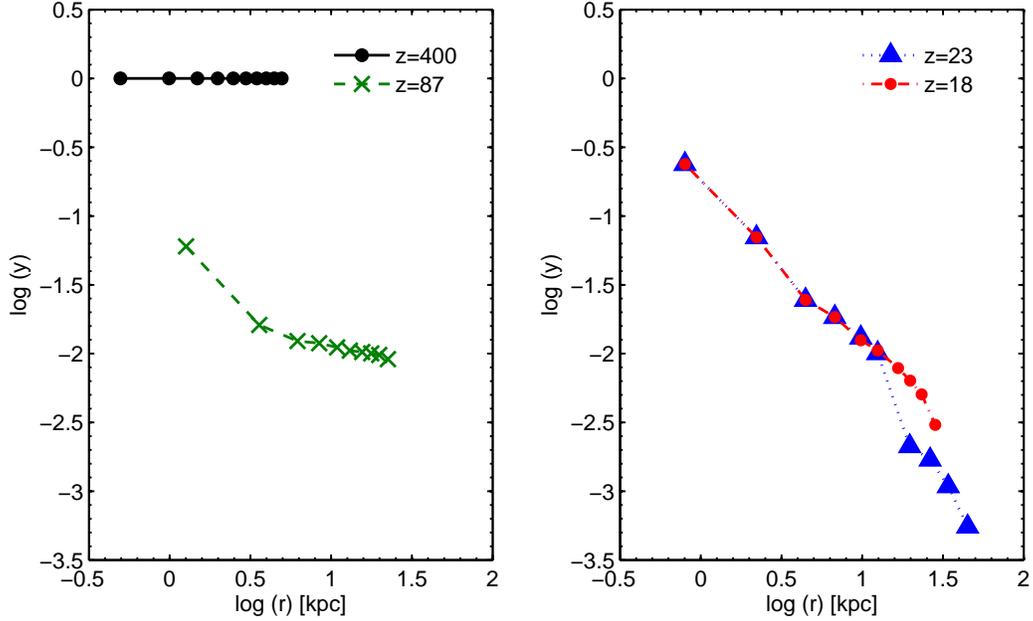}
\caption{ The evolution of density profile in logarithmic scale
normalized to the initial density $y= \rho/\rho_{initial}$, as a
function of distance from the center of structure in four different
redshifts. The initial density profile is taken top-hat distribution
with ten equidistant shells for representing the evolution of the
density. The inner most shell enters to the MOND domain at $z= 400$.
Each point on the curves notifies the position of the shell. We plot
density profile of the structure for $z=400$ and $z=87$ at the left
panel and $z=23$ and $z=18$ at the right panel for comparison. At
$z=17$ all the shells virialize and the density profile freezes with
a power-low function of $\rho \propto r^{\beta}$ where $\beta\simeq
-1.22$.} \label{dens-prof}
 \end{center}
 \end{figure}

In this part we compare the evolution of the density profile of the
structures in the spherical collapse model for the Newtonian and
MONDian regimes. In the Newtonian regime $(g>a_0)$ we have seen that
the dynamics of the shells in the structure depends only on time and
transforming to a comoving frame, the dynamics is scale independent
(see equation \ref{nc}). This means that the dynamics is invariant
under the scale transformation and the result is preserving the
initial profile of the structure. In MOND $(g<a_0)$, since the
dynamics depends on the scale unlike the Newtonian case the density
profile will change with time. In Fig. (\ref{dens-prof}) we plot the
spatial variation of density from equation (\ref{rho}) for a galaxy
mass structure in four different stages of $z = 400$, $87$, $23$ and
$18$. The initial stage at the redshift of $z = 400$ is taken a
top-hat distribution for the density when the innermost shell enters
to the MONDian regime. We continuously do Jacobian transformation
from this stage to the later times (i.e. $z<400$) and obtain the
density of each shell. While the structure evolve, the distance of
shells change with time as shown with a point on the profile
representing the position of each shell (Fig. \ref{dens-prof}). As
the inner shells enter to the MONDian regime earlier we expect the
density profile deviates from the homogeneous distribution.
Fig.(\ref{dens-prof}) shows a small deviation of the density profile
from the homogenous one at $z = 87$. At this moment all the shells
are in the MONDian domain, however as the inner shells have been
evolved faster, they are more dense than the outer shells. Each
shell after virialization freezes and preserves its density. We fit
the density profile of the structure when the outermost shell, the
latest part of the structure, virializes and the result is a
power-law function of $\rho\propto r^{\beta}$ with the index of $\beta \simeq -1.22$.\\

\subsection{Age of a structure}
A question that may be answered in this simple model is the
dependence of the age of the structure to the mass. Observations
show that small isolated structures as the globular clusters or the
center of galaxies are older than the spiral arms \cite{cha98}. In
this model we have seen that the smaller structures enter the
MONDian regime earlier, evolve faster and virialize at the higher
redshifts compared to the larger structures. As an example, in Table
(\ref{tab2}) it is shown that a structure with the mass of $10^{8}
M_\odot$ virializes at $z \simeq 71$ while the whole of galaxy
virializes at $ z \simeq 18$. The age of a structure from the
virialization up to now is given by $t_{age} = \int^{t_0}_{t_{vir}}
dt$, where $t_0$ is the present age of the Universe. Changing the
variable to the redshift results in:
\begin{equation}
t_{age} =
H_0^{-1}\int_{0}^{z_{vir}}\frac{dz}{(1+z)\sqrt{\Omega_b^{(0)}(1+z)^3
+ \Omega_r^{(0)}(1+z)^4 + (1-\Omega_t^{(0)})(1+z)^2}},
\end{equation}
where we have adopted the cosmological parameters in
(\ref{mondcos}). The age of structures depending on their masses is
given in Table (\ref{tab2}).
\subsection{Contribution of dark matter in the smaller structures}
The other question in the formation of the structures is that why
the smaller structures are mainly made of the baryonic matter and
have a less dark matter compared to the larger ones. This feature of
the structures can be explained if we can show that in addition to
the fast evolution of the smaller structures, the acceleration of
the structure in terms of $a_0$, ($g_N/a_0$) gets larger. To show
this property of the structures in MOND, we calculate the
gravitational acceleration at the virialization time of each shell
and compare it with $a_0$. As the density of a structure at this
stage changes with $\rho\propto r^{\beta}$, the gravitational
acceleration will depend on $r$ as $g_N(r) \propto r^{\beta +1}$.
For $\beta = - 1.22$ the smaller radii should have larger
acceleration compare to the larger ones.

\begin{figure}
\begin{center}
\includegraphics[angle=0,scale=0.7]{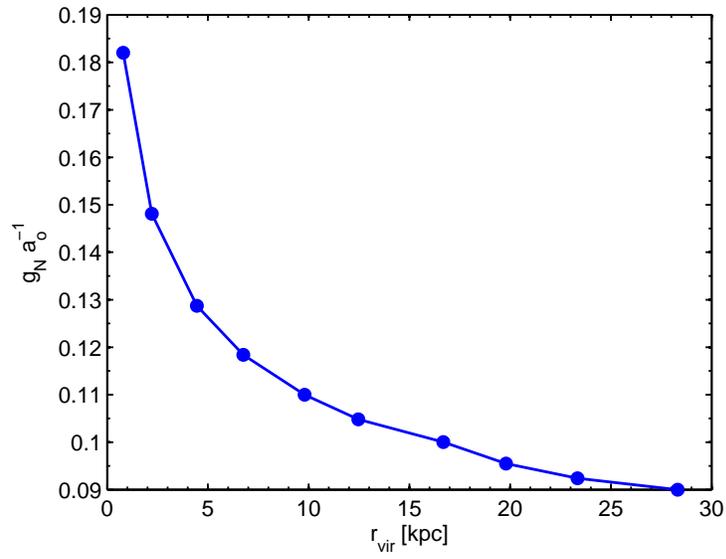}
\caption{The Newtonian gravitational acceleration of each shell
normalized to $a_0$ as a function of shell size at the virialization
time. Spots on the curve represent the acceleration and the position
of each shell. The gravitational acceleration in the inner shells is
larger than the outer shells.}\label{na}
\end{center}
\label{Newt_acceleration}
\end{figure}

\begin{figure}
\begin{center}
\includegraphics[angle=0,scale=0.7]{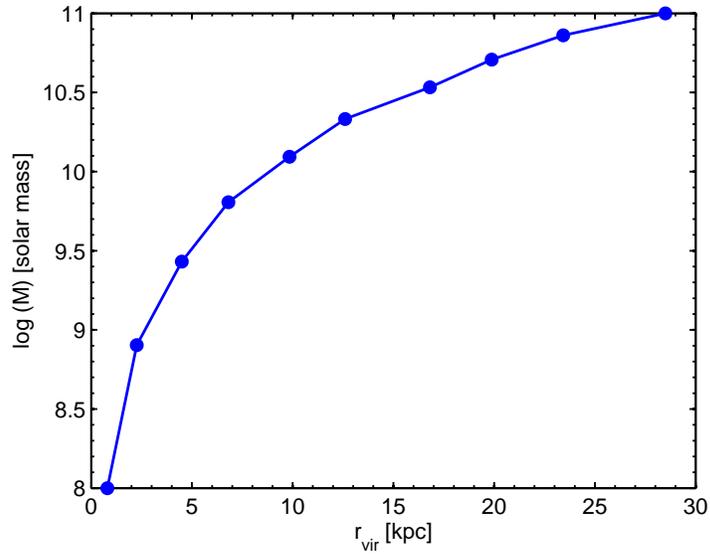}
\caption{The mass of structure in terms of size of structure after
virialization. Spots on the curve represent the the corresponding
value for each shell. Structures with the mass smaller than
$\simeq10^{10}M_{\odot}$ are virializad with the radius smaller than
$\sim10 kpc$.}\label{f7}
\end{center}
\end{figure}

In Figure (\ref{na}) we plot the gravitational acceleration for each
shell at the time of virilization in terms of the virialization
radius. For the outer shells the gravitational acceleration,
$g_N/a_0$ is smaller than the inner shells. This means that the
smaller structures after virialization need less dark matter compare
to the larger ones in the standard CDM scenario. Finally we plot the
mass of the structure after virialization in terms of its size in
Figure (\ref{f7}). This feature reveals the general property of
structures in MOND and an accurate result to compare with the
observation may be obtained by N-body simulation of the structure
formation.

\section{Conclusion}
\label{conc} Summarizing this work, we extend the spherical collapse
model to study the generic properties of the structure formation in
MOND. We showed that in the MONDian scenario, structures can evolve
without a need to the dark matter and have the following three main
features: ({\bf a}) MONDian spherical collapse unlike to that of CDM
does not preserve the initial density profile of the structure. In
MOND starting with an initial homogenous density profile the
structure virializes with a power-law distribution of the matter,
having singularity at the center. ({\bf b}) We showed that the small
scale structures enter the MONDian regime earlier than the larger
ones, evolve faster and virialize at higher redshifts. This picture
from the spherical collapse model in MOND provides a bottom-top
scenario for the structure formation in which the smaller structures
formed before the larger ones. This result is compatible with the
observations where the old stars are located in the smaller
structures such as the globular clusters or center of galaxies.
({\bf c}) Finally we showed that while the smaller structures enter
the MONDian regime before the larger ones, they have larger
gravitational acceleration at the virialization time and hence need
a less dark matter in CDM scenario.

\acknowledgments We would like to thank anonymous referee for useful
comments. Also we would like to thank M. Nouri-Zonoz for reading the
manuscript and giving useful comments.

\clearpage

\newpage

\newpage

\newpage

\newpage

\newpage

\begin{table}
\begin{center}
\caption{The parameters of each shell at the entrance time to the
MONDian regime. The first column represents the index of shell. The
shells are taken equidistant when all the structure is in the
Newtonian regime. The second column indicates the mass that
enveloped by the corresponding shell. The third column shows the
radius of each shell at the time of crossing the MOND domain and the
forth column is the corresponding redshift. The fifth column is the
density contrast at the entrance time and the sixth column shows the
ratio of the kinetic energy per mass to $W = r dV/dr$ of each shell
at the entrance time. \label{tab1}}
\begin{tabular}{crrrrr}
\tableline\tableline
i &$M_i [10^{11}M_{\odot}]$ & $r_{i}(t_{enter})[kpc]$ & $z_{eneter}$ & $\delta_{enter}\times 10^{-5}$ & ${2T^i_{ent}}/{W^i_{ent}}$\\
\tableline
1&$0.001$&$0.45$ &  $400$ & $2.71 $ & 2.24\\
2&$0.008$&$1.25$ &  $297$ &$3.70 $ & 2.28\\
3&$0.027$&$2.40$ &  $242$ & $4.52 $ & 2.32\\
4&$0.064$&$3.55$ &  $218$ & $5.03 $ & 2.36\\
5&$0.124$&$5.03$ &  $195$ & $5.61 $ & 2.41\\
6&$0.215$&$6.35$ &  $184$ & $ 5.97 $& 2.48\\
7&$0.341$&$8.17$ &  $170$ & $6.47 $ & 2.51 \\
8& $0.510$&$9.77$ &  $161$ & $6.81 $& 2.53\\
9& $0.725$&$11.44$ &  $154$ & $7.12 $& 2.55 \\
10 & $1.000$&$13.57$&  $146$ & $7.51 $& 2.57 \\
\tableline
\end{tabular}
\end{center}
\end{table}

\begin{table}
\begin{center}
\caption{ The parameters of shells in the spherical collapse model
at the maximum radius and virialization redshift. The first column
represents the index of shells. The second column indicates the mass
that enveloped by the corresponding shell. The third, forth and the
fifth columns correspond to the maximum radius, redshift and the
density contrast at that moment, respectively. The sixth, seventh
and the eighth columns  show the virialize radius, redshift and the
density contrast, respectively. The ninth column indicates the
redshift that a shell enters to the non-linear regime $(\delta>1)$
and the last column is the age of the structure after virialization.
\label{tab2}}
\begin{tabular}{crrrrrrrrrrr}
\tableline\tableline $i$
&$M[10^{11}M_\odot]$&$r_{max}[kpc]$&$z_{max}$
&$\delta_{max}$&$r_{vir} [kpc]$ &$z_{vir}$& $\delta_{vir}$&$z_{nl}$ & age[Gyr]\\
\tableline
$1$ &$0.001$ &$1.32$  &$106.52$ &$1.01$ &$0.79$& $71.46$ &$28.51$& $106.52$& 12.48\\

$2$ &$0.008$ &$3.72$  &$73.94$ &$1.25$ &$2.21$ &$49.84$ &$34.37$& $78.57$& 12.43\\

$3$ &$0.027$ &$7.43$  &$58.88$ &$1.44$ &$4.45$ &$38.52$ &$36.82$& $65.66$& 12.38\\

$4$ &$0.064$ &$11.24$ &$49.53$ &$1.52$ &$6.85$ &$32.57$ &$38.32$& $54.28$& 12.34\\

$5$ &$0.124$ &$16.24$ &$43.44$ &$1.65$ &$9.79$ &$28.41$ &$40.30$& $48.75$& 12.29\\

$6$ &$0.215$ &$20.77$ &$39.26$ &$1.72$ &$12.46$ &$25.57$ &$42.55$& $44.58$& 12.25\\

$7$ &$0.341$ &$27.72$ &$35.23$ &$1.84$ &$16.01$ &$22.80$ &$44.49$& $40.66$& 12.19\\

$8$ &$0.510$ &$32.78$ &$32.65$ &$1.89$ &$19.79$ &$21.11$ &$45.05$& $38.19$& 12.17\\

$9$ &$0.725$ &$38.63$ &$30.75$ &$1.96$ &$24.01$ &$19.83$ &$46.21$& $36.32$& 12.15\\

$10$&$1.000$ &$46.98$ &$28.02$ &$2.06$ &$28.51$ &$17.99$ &$48.72$& $33.66$& 12.09\\

\tableline
\end{tabular}
\end{center}

\end{table}

\end{document}